\documentclass[smallextended]{svjour3}       
\journalname{}
\usepackage[comma,authoryear]{natbib} 
\usepackage[utf8]{inputenc}  
\usepackage{graphicx}

\usepackage[english]{babel}
\usepackage{amsmath}
\usepackage{amssymb}
\usepackage{dsfont}
\usepackage{cases}
\usepackage[shortlabels]{enumitem}
\usepackage{color}
\usepackage{verbatim}
\usepackage{setspace}

\usepackage{tikz}

\usepackage{hyperref}

\definecolor{darkgreen}{rgb}{0.1,.8,0.1}
\definecolor{darkred}{rgb}{0.8,.1,0.1}
\definecolor{darkblue}{rgb}{0.1,.1,0.8}

\usepackage{pgfplots}
\pgfplotsset{compat=newest}

\def \1{\mathds{1}}

\newtheorem{cor}{Corollary}

\newcommand{\NI}{\zeta} 
\newcommand{\NId}{\zeta_d}

\newcommand{\sus}{S} 
\newcommand{\ifd}{I} 
\newcommand{\vac}{V} 
\newcommand{\susinit}{S_{0^-}}
\newcommand{\ifdinit}{I_{0^-}}

\newcommand{\dis}{\mathcal D}
\newcommand{\phid}{\Phi_I}
\newcommand{\Ietoile}{\ifd^*}
\newcommand{\Setoile}{\sus^*}
\newcommand{\Setoiled}{\sus^{**}}

\def\beq{\begin{equation}}
\def\eeq{\end{equation}}
\def\beqn{\begin{eqnarray}}
\def\eeqn{\end{eqnarray}}

\def \P{\mathbb{P}}

\def \R{\mathbb{R}}

\newcommand{\rv}{r_V}
\newcommand{\ri}{r_I}
\newcommand{\vpI}{\varphi_I}
\newcommand{\vpV}{\varphi_V}
\newcommand{\CI}[3]{J_{indi}(#1)} 
\newcommand{\CId}[3]{J_{indi}^{\dis}(#1)} 
\newcommand{\CIP}[3]{J_{pure}(#1)} 
\newcommand{\CIPd}[3]{J_{pure}^\dis(#1)} 
\newcommand{\si}{\theta_I}
\newcommand{\um}{u_{max}}
\newcommand{\se}{\theta} 
\newcommand{\CG}[2]{J_{soc}(#1, #2)}

\newcommand{\levelzd}{\mathcal C_{\rv / \ri}}

\newcommand{\eqfrontiere}{\ifd + (\rv / \ri) \sus +(\gamma/\beta) \ln(1-\rv/\ri)}
\newcommand{\eqfrontiereinit}{\ifdinit + (\rv / \ri) \susinit +(\gamma/\beta) \ln(1-\rv/\ri)}

\begin{document}

\title{Individual vaccination as Nash equilibrium 
in a SIR model\thanks{L.L. and G.T. acknowledge a FSMP DIM grant from the \^Ile de France region and the 
 Foundation Sciences Math\'ematiques de Paris.}}
\subtitle{with application to the 2009-10 Influenza A(H1N1) epidemic in France}
\author{Laetitia LAGUZET \and Gabriel TURINICI}
\institute{L. Laguzet ORCID : 0009-0000-9655-0880  \at
              CEREMADE, Universit\'e Paris Dauphine, PARIS 75016, FRANCE \\
              \email{laguzet@ceremade.dauphine.fr}           
           \and
           G. Turinici ORCID: 0000-0003-2713-006X \at
              CEREMADE, Universit\'e Paris Dauphine, PARIS 75016, FRANCE \\
              Tel.: +33-1-44054858\\
              \email{Gabriel.Turinici@dauphine.fr}          
							\\
							and
							\\
							Institut Universitaire de France
}
\date{\bf August 2015}
\maketitle

\begin{abstract} 

The vaccination against {ongoing} epidemics is seldom compulsory but remains one of the most classical means to fight epidemic propagation. However recent debates concerning the innocuity of vaccines and their risk with respect to the risk of the epidemic itself lead to severe vaccination campaign failures and new mass behaviors appeared driven by individual self-interest. 
Prompted by this context we analyze, in a Susceptible-Infected-Recovered (SIR) model,
whether egocentric individuals can reach an equilibrium with the rest of the society. 
Using techniques from the "Mean Field Games" theory, we extend previous results and show that an equilibrium exists and characterizes completely the individual 
best vaccination strategy (with or without discounting). 
We also compare with a strategy based only on overall societal optimization and exhibit
a situation with non-negative price of anarchy. Finally, we apply the theory to the 2009-2010 Influenza A (H1N1) vaccination campaign in France and hint that a group of
 individuals stopped vaccinating at levels that indicated a pessimistic perception of the risk of the vaccine.



\end{abstract}

\section{Introduction}  \label{sec:introduction}

The vaccination, when available, is one of the most classical defense against an evolving epidemic. Theoretical works have been proposed to describe the optimal vaccination policy as function of the vaccine cost and epidemic dynamics and severity, 
see for instance~\citep{Hethcote73,abakuks1974opt,morton74,sethi78,heesterbeek,Laguzet2015180}.
The point of view  of these initial studies is that of a benevolent planner that optimizes the overall societal welfare. However in most situations vaccination is not compulsory and people have the choice to vaccinate or not. For a number of childhood diseases (with no compulsory vaccination) a decay in the vaccination coverage has been observed in developed countries and debates over the danger of vaccines and their usefulness emerged.
This suggested that the individual point of view has also to be taken into account: any individual may choose to vaccinate or not and the 
epidemic dynamics is influenced by all these individual choices. 
The individual defines a cost $r_V$ incurred if he takes the vaccine 
and a cost $r_I$ incurred if infected; the costs can be expressed in terms of money, medical side effects or general morbidity. For a
discussion on this topic refer to~\citet{qaly1,daly1,qalydaly06} and related literature on QALY/DALY measuring scales.
The cost $r_I$ is to be weighted by the probability to be infected; it is immediate that when the epidemic is near extinction there 
is very low incentive to vaccinate (because the probability to be infected is small). Thus people stop vaccinating before the epidemic stops, which can cause in fact the epidemic to start again. In theory the situation can be highly unstable oscillating between no vaccination, followed by an epidemic outburst, then massive vaccination followed by epidemic near extinction and so on.
An important question is the existence of a stable equilibrium and the impact on the vaccination coverage. 

Several works on this topic appeared as early as \cite{FINE01121986}, 
\cite{Brito199169} and \cite{geoffard1997}
and ask the question of disease eradication, market equilibrium and externalities in relationship with vaccination;
recently \citet{Bauch02092003} (using a SEIR model) contributed to the renewal of interest on vaccination policies and individual 
decisions. In a subsequent paper,~\citet{bauch2004vac}
use a SIR model with vital dynamics to describe the propagation of a childhood disease; the individual choices converge to a sub-optimal vaccine coverage. The vaccination strategy of the individual is of all-or-nothing type and time-independent.
Furthermore,~\citet{bauch2005imi} proposes a learning process based on a "rule of thumb"  to explain why and how the people vaccinate, 
resulting in a time-dependent vaccination strategy. The presence of a time-dependent optimal strategy is more realistic and an advance over the previous work but the model is dependent on the "rule of thumb" chosen {\it a priori}.
This study was completed by~\citet{shim2012the}
where the depart from an $100\%$ egocentric decision is investigated.
\citet{behavior2} and~\citet{galvani07} also model the vaccination behavior and apply the results to the yellow fever scare in Brazil. \citet{Buonomo20089} also introduce a feed-back mechanism but for a SEIR system.

In a very elegant approach, \citet{Francis20042037} studied a situation that match very well our setting: the equilibrium in the SIR model.
They study in addition the impact of taxes and subsidies and propose revenue-neutral health policies to encourage vaccination. Although the techniques are very intuitive, they do not explicitly introduce a cost functional for the individual and as such this approach has 
yet to be extended to general settings (for instance when using discounting). 
We present in Section~\ref{sec:discounted} additional comments comparing the
two approaches.

In another work~\citet{reluga2006evo} studied the impact of the perceptions of the relative risk between vaccine and 
epidemic; they used a SIR model with vital dynamics and also discussed the imitation dynamics.
\citet{Galvani27032007} consider a double SIR periodic model of influenza with vaccination and 
two age groups (more or less than $65$ years). Vaccination is separated from dynamics and arrives once at the beginning of each season. The effect of the 
group dependent vaccination is analyzed. \citet{cojocaru2} and~\citet{cojocaru3,chen06SI} also consider the mathematical questions related to the presence of several groups having distinct epidemic characteristics.

Further contributions for models with vital dynamics include~\citet{dOnofrio07,dOnofrio08,Reluga201167}.

A review on the relationship between human behavior and epidemic dynamics is proposed by~\citet{salathe}. In particular they discuss 
the relationship between the timescales of the vaccination and epidemic dynamics, which is also the object of several
contributions by~\citet{vardavas07,breban07}. \citet{Chen2009125} discusses how the available information change the decisions made by
an individual.

Even if realistic vaccination behavior is very likely to depend on the imitation and altruistic dynamics, there is a need to separate this
part from the  optimization of the cost-benefit ratio at the individual level.

{A recent example of low individual vaccination is}
the 2009-10 influenza A (H1N1) epidemic. 
In this case the vaccine is only relevant for {one} epidemic season and therefore {previous} 
models do not describe accurately the dynamics.
We compare in 
Table~\ref{table:target} the difference between the target coverage of the vaccines, as defined by the sanitary authorities, and the 
effective rate of vaccination obtained. In many European countries one order of magnitude separates the two.

\begin{table}[ht] 
\centering
\begin{tabular}{|c|c|c|}
\hline
Country & Target  & Effective rate \\
 & coverage &of vaccination \\
\hline
Germany & $30-100\%$ & $8-10\%$ \\
\hline
Spain & $40\%$ &$4-27.1\%$ \\
\hline
France & $70-75\%$ & $7.9-8.5\%$ \\
\hline
Italy & $40\%$ & $1.4-4\%$ \\
\hline
\end{tabular}
\caption{Vaccination Coverage expected and realized in different countries as
percentage of population during the 2009-10 Influenza A (H1N1) epidemic. 
Sources: \citet{couverturefrance,coverageeurope,vaccinationAllemagne1,vaccinationFrance1,determinantsvaccination},\citep[page 157]{rapportparlementaire}.
}
\label{table:target}
\end{table}

But today there is still need for theoretical guidance to explain this data 
{with existing models}.
{In particular}
we expect the optimal vaccination strategy to be time-dependent.
 We also want to 
 identify the vaccination dynamics and propose a model that only takes into account the effect of the individual cost optimization; such a result can then help identify the  impact of other effects: imitation, altruism.

We prove the existence of an equilibrium between individual vaccination and an epidemic propagation described by a deterministic SIR-model
(with or without discounting). We compare the theoretical result with the optimal policy at the societal level and describe the differences between the two. Finally we apply the theoretical result to the Influenza A (H1N1) 2009-10 epidemic season in France and observe that people were not homogeneous in their perception of the $r_V/r_I$ quotient: while some saw the vaccine as harmless, a non-negligible portion of the population saw the vaccine as potentially risky.

\subsection{The model}  \label{sec:themodel}

We consider an epidemic spreading in a non-immune population; the dynamics of the epidemic follows a 
SIR model~\citep{Anders2010,heesterbeek}. 
The epidemic is supposed to take place at a rapid time frame (several months up to $1$-$2$ years) when compared with the demographic dynamics (births, deaths) and therefore the model does not take into account any vital dynamics.

The costs incurred by an infected individual (either in terms of monetary value or of medical condition) are the same for any individual and are denoted by $r_I$. 
We also suppose that a vaccine giving lifelong immunity exists. Its cost 
 $r_V$
 is considered known by all the individuals
and takes into account not only the economic price but also all possible side-effects of the vaccine.
The people in the Susceptible class can choose to vaccinate or not (people in other classes cannot vaccinate).

The mathematical equation describing the SIR model with vaccination is:
\begin{equation}
\begin{cases}
    d\sus(t) = - \beta \sus(t) \ifd(t)dt - dU(t), &   \sus(0^-) = \susinit,\\
    d\ifd(t) = \Big(\beta \sus(t) \ifd(t) - \gamma \ifd(t)\Big)dt, &  \ifd(0^-) = \ifdinit.		\\ 
\end{cases}
\label{sys:SIRU}
\end{equation}
Here 
$dU(t)$ is the vaccination rate,
$\sus(t)$ is the proportion of individuals in the Susceptible class and $\ifd(t)$ is the proportion in the Infected class.
The proportion of people in the Recovered class is $\int_{0^-}^t  \gamma \ifd(\tau) d\tau$
and
 $\vac(t) := \int_{0^-}^t dU(\tau)$ is the proportion of people that vaccinated by the time $t$.
The parameters $\beta$, $\gamma$ and the initial conditions $\susinit$ and 
 $\ifdinit$ 
are supposed known.
Note that when vaccination is instantaneous the function $\sus(t)$ may be discontinuous and its derivative 
is a measure because $dU(t)$ only exists as a measure. 
We refer to the Appendix~\ref{appendix:sirequationmeasure} for the mathematical details.

Equations~\eqref{sys:SIRU} only represents the overall, societal dynamics.  At the microscopic level, the individual dynamics is 
modeled by a continuous time Markov chain with individual jumping between the 
Susceptible, Infected, Recovered and Vaccinated classes, as illustrated in Figure~\ref{fig:individualMarkov}. 
\begin{figure}[!ht]
\centering
\includegraphics[width=0.8\textwidth]{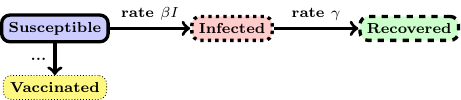}
\caption{Individual dynamics: continuous time Markov jumps between Susceptible, Infected, Recovered and  Vaccinated classes.}
\label{fig:individualMarkov}
\end{figure} 

The jump from the Susceptible to the Vaccinated class depends on the willingness of the individual to vaccinate.
We suppose that each individual is aware of the propagation dynamics at the societal level~\eqref{sys:SIRU} but cannot influence it. His only concern is whether it is useful for himself to vaccinate or not and when to do it. 
This decision is called a {\it strategy}. A first interpretation of the strategy is the instant $t\in [0,\infty[$ when the individual vaccinates (if he is still in the Susceptible class at time $t$). Such a strategy is denoted $\Pi_t$; if  the individual never vaccinates
his strategy is $\Pi_\infty$.
However it turns out (see discussion in Appendix~\ref{appendix:individualstrat}) that 
it is  better to represent the individual strategies as {\it mixed strategies} which are probability laws on the set of all pure strategies.
We represent such a probability law by its CDF (cumulative distribution function) $\vpV$ with
$\vpV(0^-)=0$; then
$\vpV(t)$ represents the probability to choose a pure strategy in the interval $[0,t]$. When
$\vpV(\infty) \neq 1$ this means that with probability
$1-\vpV(\infty)$ the individual never vaccinates. When  $\vpV(0)>0$ then the individual vaccinates immediately with probability
$\vpV(0)- \vpV(0^-)= \vpV(0)$.
{When the vaccination capacity is limited by a constant $\um>0$, the CDF $\vpV$ has to comply with the constraints given in equation~\eqref{eq:constraint0}.}

Another useful mathematical object is the cumulative probability of infection in $[0,t]$, denoted  $\vpI(t)$:
\begin{equation} \label{eq:vpI_expli}
\vpI(t) = 1- e^{- \int_{0^-}^t \beta \ifd ( \tau) d \tau}.
\end{equation} 

Note that $\vpI(t)$ depends on $\susinit$, $\ifdinit$ and $U$; however to ease notations this dependence will not be written explicitly.

Given the epidemic dynamics (fully determined by $\susinit$, $\ifdinit$ and $U$) 
the individual can associate a cost, denoted $\CIP{\cdot}{\susinit, \ifdinit}{U}$, to each pure strategy. For instance $\CIP{\Pi_0}{\susinit, \ifdinit}{U}=r_V$.
For general $t\in[0,\infty[$  the cost of $\Pi_t$ is the sum 
of the cost of infection $r_I$ weighted by the probability $\vpI(t)$ that infection occurs before time $t$ plus the  
cost of vaccination weighted by the probability $1-\vpI(t)$ that infection did not occurred before time $t$:
\begin{equation}
\CIP{\Pi_t}{\susinit, \ifdinit}{U} = \ri \vpI(t)  + \rv (1 - \vpI(t)).
\label{eq:costpuret}
\end{equation}
The non-vaccination strategy $\Pi_\infty$ costs $r_I \vpI(\infty)$. Summing up the terms, the cost of the mixed strategy with CDF $\vpV(t)$, denoted $\CI{\vpV}{\susinit, \ifdinit}{U}$, is:
\begin{eqnarray}
& \ & \CI{\vpV}{\susinit, \ifdinit}{U} = 
(1- \vpV(\infty))\CIP{\Pi_\infty}{\susinit, \ifdinit}{U}
+
\int_{0^-}^{\infty} \CIP{\Pi_t}{\susinit, \ifdinit}{U} d\vpV(t) 
\nonumber \\ & \ & 
=\ri \vpI(\infty) + \int_{0^-}^{\infty} \Big[
\rv - \ri \vpI(\infty) + (\ri - \rv) \vpI(t) \Big] d\vpV(t) .
\label{eq:ci2}
\end{eqnarray}

Alternative equivalent definitions have been proposed in the general context of {\it timing games}, see~\cite[page 118]{drewtirole}.

When a discount factor $\dis > 0$ is introduced, the discounted risk of infection is:
\begin{equation}  
\phid (t)  = \int_{0^-}^t e^{-\dis \tau} d \vpI(\tau), \ \phid (0^-)=0.
\label{eq:phid}
\end{equation}

The cost of pure strategies is for $t\in[0,\infty[$:
\begin{equation}
\CIPd{\Pi_t}{\susinit, \ifdinit}{U} = \ri \phid(t)  + e^{-\dis t}\rv (1 - \vpI(t)).
\label{eq:costpuretd}
\end{equation}
In this case  $\CIPd{\Pi_\infty}{\susinit, \ifdinit}{U}= r_I \phid(\infty)$. The 
the cost of the mixed strategy with CDF $\vpV(t)$ is:
\begin{equation}
\CId{\vpV}{\susinit, \ifdinit}{U} = \ri \phid(\infty) + \int_{0^-}^{\infty} \Big[ \ri (\phid(t) - \phid (\infty) ) + \rv e^{- \dis t} (1-\vpI(t)) \Big] d\vpV(t).
\label{eq:cid2}
\end{equation}

We work under the constraints $(S,I) \in \overline{\Omega}$, where
\begin{eqnarray}
\Omega = \{ (\sus, \ifd) \in \R^2  \ | \  \sus, \ifd > 0, \sus + \ifd < 1\}.
\label{eq:defomega}
\end{eqnarray}

{

Recall that the global dynamics is an aggregation of individual dynamics. Therefore 
 $\ifd(t)$ in Equation~\eqref{sys:SIRU} is the same 
as in~\eqref{eq:vpI}
and
a compatibility relation has to exist between $dU(t)$ (societal vaccination) and $\vpV$ (individual vaccination).

To make explicit this last compatibility requirement one  has to investigate the relationship between the Markov chain of one individual and the evolution dynamics of the population; such an endeavor is beyond the scope of this work but has been explored in several papers 
starting with~\citet{kurtz70}, see also~\citet[Theorem 10.2.3 pages 373-374]{sandholm10}  for an application to population dynamics. Further results very close to the setting of this work are given by~\citet{gomesmfg,gueantmfg}. The conclusion is that the societal dynamics 
is compatible with the individual dynamics when:
\begin{equation}
dU(t) = \frac{d \vpV(t)}{1-\vpV(t)} \sus(t).
\label{eq:UGequilibre} 
\end{equation}

An intuitive understanding of this formula is the following: for a given individual, the probability that it 
is vaccinated at time $t+\Delta t$ provided that it was not vaccinated at time $t$ equals $\frac{\vpV(t+\Delta t) - \vpV(t)}{1-\vpV(t)} + o(\Delta t)$ (see Equation~\eqref{eq:vpid2} for a similar computation). At the same time, since individuals are all the same, the same probability has to be $\frac{U(t+\Delta t)-U(t)}{\sus(t)} + o(\Delta t)$. Comparing the two in the limit $\Delta t \to 0$ we obtain the Equation~\eqref{eq:UGequilibre}.

}
\subsection{Questions regarding the equilibrium} \label{sec:questionseq}

When everybody is vaccinating this collective behavior stops the epidemic. For a given individual that still has to make 
its own choice,
the perceived benefit of vaccination is low because the risk associated with the epidemic is very low. 
Why should he vaccinate then ? The individual will therefore {\bf not}  act as everybody else.

On the contrary, when nobody vaccinates and the individual foresees a severe epidemic (because of lack of vaccination for instance), the individual is lead to vaccinate, therefore will {\bf not} act as everybody else.

In both situations there is an incoherence between the overall, societal, dynamics and the individual self-interest. A legitimate question is whether an equilibrium scenario exists where the best vaccination policy of the individual and the vaccination policy of everybody else agrees.

To illustrate this question consider the Figure~\ref{fig:Cout_fct_SG_SI}
 corresponding to the model in Section~\ref{sec:themodel}. Vaccination is represented by a parameter indicated the time during which vaccination takes place. A low time indicates low vaccination level and a large time a high vaccination level. 
We plot a three-dimensional surface of the cost $\CI{\theta_I}{\susinit, \ifdinit}{\theta_G}$ incurred by an individual which has the vaccination time 
$\theta_I$  in a circumstance when the societal vaccination time is $\theta_G$. 
Note that, with an abuse of notation, we write $\theta_G$ instead of $dU=\mathds{1}_{[0,\theta_G]}\um dt$
and $\theta_I$ instead of the solution of $d \vpV(t)=(1-\vpV(t)) \mathds{1}_{[0,\theta_I]} \um / \sus(t)  dt$, see Appendices~\ref{appendix:sirequationmeasure} and~\ref{appendix:individualstrat}.

For any societal vaccination strategy represented by the parameter $\theta_G$ the individual 
minimizes  $\theta_I \mapsto \CI{\theta_I}{\susinit, \ifdinit}{\theta_G}$. The optimal value $\theta_I^{opt}$ 
of $\theta_I$ as a function of $\theta_G$ is represented
as the solid bottom curve in the Figure~\ref{fig:Cout_fct_SG_SI}. 
The existence and uniqueness of the function $\theta_I^{opt}(\theta_G)$ (for general societal vaccination strategies $dU$, not necessarily given in terms of vaccination time) is 
discussed 
 in Section~\ref{sec:individualtheory} and Appendix~\ref{appendix:indoptstrat} for $\dis=0$
and 
Section~\ref{sec:discounted} and Appendix~\ref{appendix:equilibre_discount} for $\dis> 0$.
As expected, for low $\theta_G$ the curve 
$\theta_I^{opt}(\theta_G)$ takes a large value while for large values of $\theta_G$ the curve $\theta_I^{opt}(\theta_G)$ 
takes small values. The equilibrium is when $\theta_I^{opt}(\theta_G)=\theta_G$.
 We therefore ask for the existence of a common point of the solid and dashed curves. The 
existence and uniqueness of an equilibrium is
discussed 
 in Section~\ref{sec:equilibrium} and Appendix~\ref{appendix:equilibre} for $\dis=0$
and 
Section~\ref{sec:discounted} and Appendix~\ref{appendix:equilibre_discount} for $\dis> 0$.
Such questions have been given a firm mathematical ground since the introduction of the "Mean Field Games" 
theory in the pioneering works of~\citet{mfgcras1,mfgcras2,mfgchm1,mfgchm2} { (see also~\citet{gomesmfg,gueantmfg} for related applications).}

\begin{figure}[!ht] \sidecaption
\centering
\includegraphics[width=0.50\textwidth]{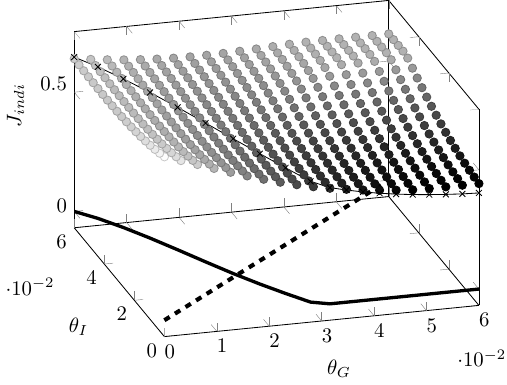}
\caption{$\CI{\theta_I}{0.62501,0.15001}{\theta_G}$ for parameters $\beta = 73$, $\gamma = 36.5$, $u_{max} = 10$, $r_I =1 $, $r_V = 0.5$ and $\dis = 0$. The solid bottom curve represent the optimal individual choice $\theta_I^{opt}(\theta_G)$  and the dashed bottom curve is $\theta_G = \theta_I$.}
\label{fig:Cout_fct_SG_SI}
\end{figure} 

\section{Illustration of the theoretical results: the undiscounted case}

We consider in this Section the undiscounted case corresponding to $\dis=0$.

\subsection{Individual optimal strategy for arbitrary epidemic propagation} \label{sec:individualtheory}

If the societal vaccination strategy $dU$ is given, there exists an individual strategy $\vpV$ that minimizes 
the individual cost~\eqref{eq:ci2}. The strategy is unique (except degenerate cases). The technical details concerning the 
existence, uniqueness and other properties of the optimal individual strategy are proved in the Appendix~\ref{appendix:indoptstrat}. 
Here we only illustrate these results in a particular case.

The theoretical results tell us that, in order to find the optimal decision at time $t$, an individual has to calculate the quantity 
$\frac{\vpI(\infty)- \vpI(t)}{1- \vpI(t)}$
 and compare it with $\rv/\ri$. While 
$\frac{\vpI(\infty)- \vpI(t)}{1- \vpI(t)} \ge \rv/\ri$ it is optimal for
the individual to vaccinate, otherwise he should not vaccinate. Note that 
$\frac{\vpI(\infty)- \vpI(t)}{1- \vpI(t)}$
can only decrease during the course of an epidemic.

The Figure \ref{fig:traj_strat_indi} gives an example of such a situation with $U=0$.  Until the time $0.06$, 
$\frac{\vpI(\infty)- \vpI(t)}{1- \vpI(t)} \ge \rv/\ri$  
and the individual vaccinates. After this time individual vaccination stops. 
The probability to be vaccinated is constant equal to $0.07\%$ after time $0.06$.

\begin{figure}[!ht]
\centering
\includegraphics[width=0.75\textwidth]{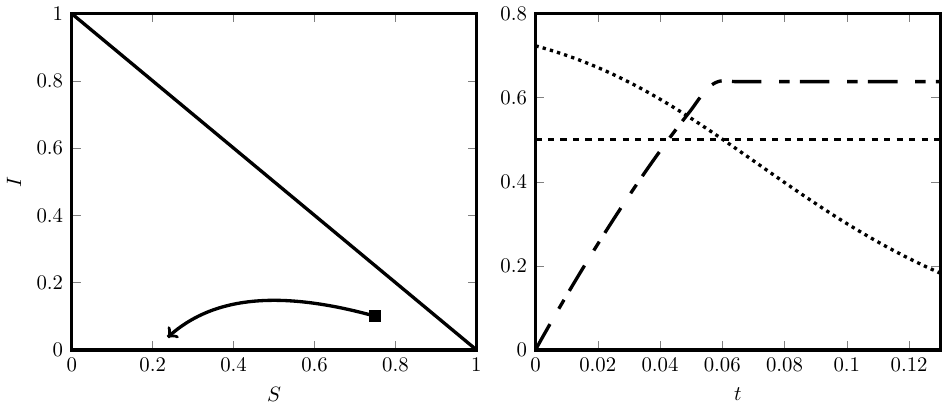}
\caption{The parameters used are $(\susinit, \ifdinit) = (0.75,0.1)$, $\beta = 73$, $\gamma = 36.5$, $r_I =1 $, $r_V = 0.5$,
$\um=10$ and constraint in Equation~\eqref{eq:constraint0}.
{\bf Left:} The trajectory 
$(\sus(t),\ifd(t))$ of the system \eqref{sys:SIRU} with $U=0$. {\bf Right:} The decreasing dotted curve is 
$t \mapsto \frac{\vpI(\infty)- \vpI(t)}{1- \vpI(t)}$, 
the constant densely dashed line is the level $\rv/\ri$ and the increasing 
dashed-dotted curve is the  probability of vaccination over time.}
\label{fig:traj_strat_indi}
\end{figure}

\subsection{Equilibrium strategies} \label{sec:equilibrium}

In classical settings that disregard individual choices (see~\citep{abakuks1974opt,Laguzet2015180}) 
the cost for the society defined as:
\begin{equation}
\CG{\susinit, \ifdinit}{U} = \int_{0^-}^{\infty} \ri \sus(t)  \frac{d \vpI(t)}{1-\vpI(t)} + \rv dU(t),
\label{eq:CG}
\end{equation}
is the only quantity of interest and is minimized.
An optimal societal strategy can be found, which will be denoted from now on (OS). 
The strategy (OS) partitions the domain $\Omega$ into a region of vaccination and a region without vaccination. 
The 
frontier of the two regions is the curve
$
\left\{ (\sus,\ifd) \in \Omega \left| \partial_{\sus}\NI(\sus,\ifd)= r_V/r_I \right. \right\}
$.

\begin{remark}
We recall that for any $(\sus, \ifd) \in \Omega$ one can introduce the number $\NI(\sus, \ifd)$ of infected people (in absence of vaccination) for a trajectory starting
 in $(\sus,\ifd)$. In particular $\NI$ is the solution of the following equation 
(see~\citep{abakuks1974opt,theseAbakuks,Laguzet2015180}):
\begin{equation}
1 - \NI(\sus, \ifd) / \sus = e^{ - \frac{\beta}{\gamma} (\ifd + \NI(\sus, \ifd))}.
\label{eq:NI}
\end{equation}
\end{remark}

In our setting, the situation is not a simple optimization. Any individual optimizes its cost but the coherence of the model
{\bf requires} that the aggregation of individual optimal policies $\vpV$ result in the global vaccination policy $dU$. Therefore
the situation is more adequately described by an equilibrium in the form of a fixed point property: 
a global vaccination $dU$ gives optimal 
individual responses $\vpV$ which sum up to form a societal response $dU'$. The equilibrium is reached only when $dU=dU'$.

The first good news is
that the results in Appendix~\ref{appendix:equilibre} show that an equilibrium 
always exists, i.e., even if the individuals are only
driven by self-interest some vaccination level is conserved (and does not drop to zero). This is completely coherent with other results from the literature, see~\citet{bauch2004vac} and related works.

The couple of optimal individual strategy and resulting societal strategy in equilibrium will be denoted from now on (EIS).
The equilibrium (EIS) is also described as a partition of 
$\Omega$ into a region of vaccination and a region without vaccination. 
Individuals will vaccinate in the first region and stop vaccination upon reaching the boundary of the second region.
The line that delimits the two regions is of equation:
\beq
\left\{(\sus,\ifd) \in \Omega \left| \eqfrontiere =0 \right. \right\}.
\eeq

The Proposition \ref{ppt:emboitement_region} shows that the vaccination region of the (OS) strategy 
includes strictly the vaccination region of the (EIS) strategy.

The Figure~\ref{fig:3regions} presents the comparison of the regions in $\Omega$. The response of an individual to a given state $(\sus, \ifd)$ of the epidemic dynamics
depends on the location of $(\sus, \ifd)$ in $\Omega$.
In the gray region  the individual (in the (EIS) strategy) 
will vaccinate and also the (OG) strategy is to vaccinate. 
In the hashed region the individual (in the (EIS) strategy) 
will not vaccinate but the (OG) strategy is to vaccinate. 
In the white region the individual (in the (EIS) strategy) 
will not vaccinate and the (OS) strategy is to not vaccinate. 

Therefore only the hashed region, delimited by the curve 
$\partial_{\sus}\NI(\sus,\ifd)= r_V/r_I$
and the line
$\eqfrontiere = 0$, is conflictual, in the sense that the individual does not have incentive to vaccinate but the 
 societal best decision is to vaccinate.
In this region the individual will tend to \textit{free-ride} on the vaccination of the rest of the society.

\begin{figure}[!ht] \sidecaption
\centering
\includegraphics[width=0.60\textwidth]{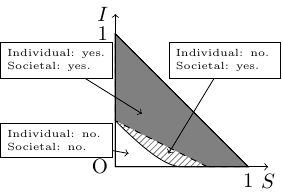}
\caption{The representation of the domain $\Omega$  and its
 regions.
Gray region: vaccination in the  (OG) strategy 
and in the (EIS) strategy.
Hashed region: vaccination in the (OG) strategy 
but not in the (EIS) strategy.
White region: no vaccination in the (OG) strategy 
and no vaccination in the (EIS) strategy.
}
\label{fig:3regions}
\end{figure} 

We illustrate in Figure \ref{fig:3regions_traj} the societal trajectories for three possible strategies in 
equilibrium, that is, satisfying the constraint~\eqref{eq:UGequilibre} (but not all optimal at the individual level). 
The dashed curve with no label corresponds to no vaccination and the individual cost is $0.65$. The second dashed curve 
corresponds to the (EIS) strategy: between point $(\susinit, \ifdinit)$ and $A$, the individual vaccinates ($d\vpV =  \um / \sus(t) (1 - \vpV(t))$) and after point A, there is no vaccination $(d\vpV = 0)$. The individual cost is $0.55$. 
The third curve corresponds to the trajectory with vaccination coherent with the (OS) societal strategy: vaccination occurs between point $(\susinit, \ifdinit)$ and B. 
The individual cost is $0.53$.

A very counter-intuitive fact is that the individual cost for the (OS) strategy is {\bf lower} than 
the cost of the (EIS) strategy. Is this not in contradiction with the fact that the (EIS) strategy is the optimal strategy for an individual~? How can be something better than the optimal ? In fact the equilibrium is a Nash equilibrium which is not globally optimal.
Although everybody would be better off adopting the (OS) strategy this choice is not stable at the individual level. A new individual added to such a population has the incentive to vaccinate less, driving the equilibrium towards the (EIS) stable point.
See also Appendix~\ref{appendix:cout_moyen} for the relation between the two costs. In game theory such a circumstance is 
termed "price of anarchy".

\begin{figure}[!ht] \sidecaption
\centering
\includegraphics[width=0.60\textwidth]{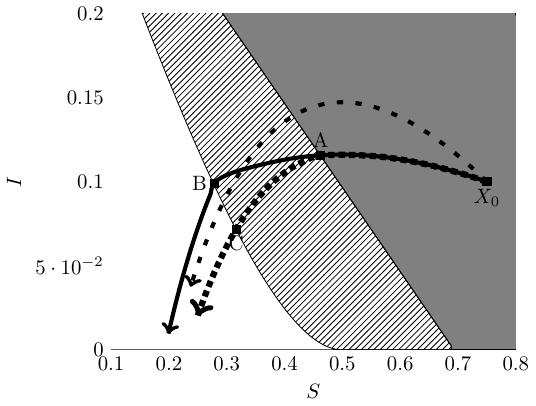}
\caption{
The dashed curve (without the point A) is with no vaccination ($dU =0$) and has individual cost $0.68$; 
the solid curve is with (OS) strategy and has individual cost $0.51$;
 the dashed curve (with point A) is with the (EIS) strategy and has individual cost $0.54$.
The parameters used for the three trajectories are $\beta = 73$, $\gamma = 36.5$, $\um = 10$, $\rv = 0.5$, $\ri =1$,
$\susinit=0.75$, $\ifdinit=0.1$. 
}
\label{fig:3regions_traj}
\end{figure}

\section{Illustration of the theoretical results: the discounted case} \label{sec:discounted}

We consider in this Section $\dis>0$. { Such a situation appears when individuals favor the present more than the future: 
when faced with two identical events, one in $t>0$ years and one in the present, the event in the future is seen as less severe (or less beneficial) than the event in the future. The attenuation factor is by definition $e^{-\dis t}$. There is no general prescription on
the precise numerical value to use for $\dis$ because it is specific to the subjective behavior of the individuals in relationship with 
a given epidemic. Discounting has been also discussed in the context of the QALY/DALY scales (see~\citet{qaly1,daly1,qalydaly06}).
}

This { situation} features several important novelties. First, the optimal individual strategies are not based on the comparison
between the infection risk and the vaccine cost. This approach, used profitably by~\citet{Francis20042037} 
(see also~\citet{Francis1997383}) is only valid when there are precisely two regions (a vaccination and a non-vaccination region) and the only pure strategies ever used are $\Pi_0$ and $\Pi_\infty$. Or, in general, it is impossible to know in advance the structure of the solution and therefore their approach needs to be extended, as shown in this work.

When $\dis >0$, vaccinating in the future makes sense 
when $\ifdinit$ is small and $\susinit$ large because the infection risk is temporarily attenuated by the 
discount factor. The criterion is the
difference between the quantity $\rv (1- e^{- \dis \Delta t})$ (gained from postponing vaccination $\Delta t$ units of time) and 
the increase in the risk of infection during the $\Delta t$ units of time. The correct solution requires proper consideration of the probability law $\vpV$, see Appendix~\ref{appendix:equilibre_discount} for details.
A graphical illustration of this situation is given in Figure~\ref{fig:traj_strat_indi_discount}.

\begin{figure}[!ht]
\centering
\includegraphics[width=0.99\textwidth]{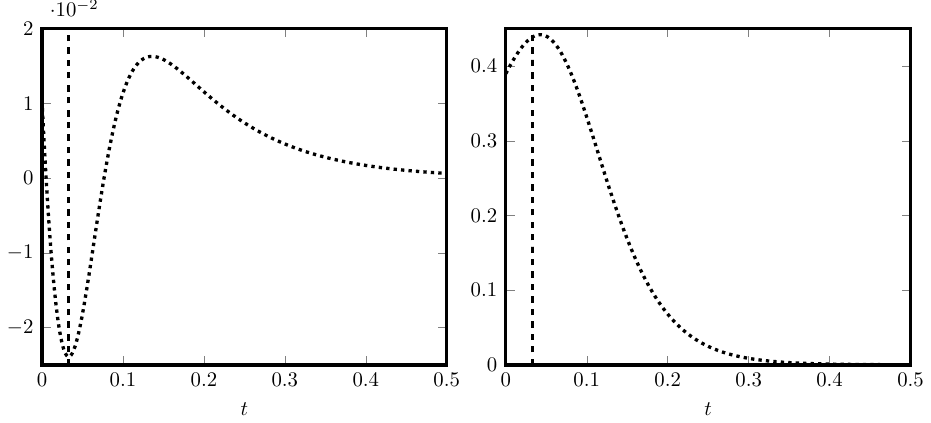}
\caption{The parameters used are $(\susinit, \ifdinit) =  (0.9, 0.04)$, $\beta = 73$, $\gamma = 36.5$, $\rv = 0.4$, $\ri = 1$ and $\dis = 10$. 
\textbf{Left:} Evolution of the function $g$ (see Appendix~\ref{appendix:equilibre_discount}). The 
optimal individual vaccination occurs at time $\tau_* = 0.0329$ (minimum of the function).
\textbf{Right:} The risk of infection 
$t \mapsto \frac{\ri (\phid(\infty) - \phid (t) )}{\rv e^{- \dis t} (1-\vpI(t))}$; the initial value is $0.3897$. 
The vaccination does not occur when the risk of infection reaches $\rv$ for the first time (at time $t= 0.0048$)
nor at the maximum value of the infection risk (time $t=0.0425$).
}
\label{fig:traj_strat_indi_discount}
\end{figure} 

When $\dis >0$ the equilibrium features three types of regions: a no-vaccination region $\Omega_n$, a instantaneous vaccination region
$\Omega_i$ and a delayed vaccination region $\Omega_d$. The zone $\Omega_d$ does not exist when $\dis=0$ or $\dis >0$ and  $\rv/\ri$ is large; however $\Omega_d$ appears 
in the realistic situation when $\dis >0$ and $\rv/\ri$ is small. The Figure~\ref{fig:regions_discount} illustrates this partition.

When the initial point $(\susinit,\ifdinit) \in \Omega_d$ the equilibrium is such that vaccination is not stable and will not be
requested until the infection level $\ifd=\Ietoile$ has been reached 
(see Appendix~\ref{appendix:zetad} for the definition of $\Ietoile$).

\begin{figure}[!ht] 
\sidecaption
\centering
\includegraphics[width=0.55\textwidth]{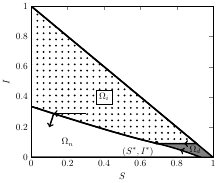}
\caption{The partition of the domain $\Omega$ in three regions: $\Omega_i$ (dotted, instantaneous vaccination), 
$\Omega_n$ (white,  no vaccination), $\Omega_d$ (gray, delayed vaccination) for 
$\gamma=36.5$,$\beta=73$, $\rv/\ri=0.4$ and $\dis =10$.
In this case  $\Omega_d \neq \emptyset$. The equilibrium dynamics is as follows: 
if $(\susinit,\ifdinit)\in \Omega_n$ it will always remains there;
if $(\susinit,\ifdinit)\in \Omega_i$ partial vaccination will bring it to $\Omega_n$ and 
total vaccination to the boundary $\sus=0$;
if $(\susinit,\ifdinit)\in \Omega_d$ it will evolve to some point $(\sus,\Ietoile)$, then vaccinate till
the point $(\Setoile,\Ietoile)$ on the boundary of all three regions, then will evolve in the region $\Omega_n$.
If on the contrary $\rv/\ri=0.6$ we have $\Omega_d = \emptyset$; this situation is not
illustrated here as it is similar to Figure~\ref{fig:3regions}).
}
\label{fig:regions_discount}
\end{figure} 

\section{An application to the Influenza A (H1N1) 2009/2010 epidemic in France} \label{sec:application}

We apply in this Section the previous theoretical results to a practical situation. Recall that in our model individuals are all the same and 
are rational optimizing agents with perfect knowledge and foresight. Although this is never true in reality, the model 
may still prove useful in order 
to understand in what regards the reality is different from the model. In particular we will show here that 
it is unlikely that all individuals shared the same $\rv$ in the specific example chosen in this Section.

\subsection{Epidemic context}

The  2009/2010 Influenza A (H1N1) epidemic in France is a recent example of 
vaccination campaign that displays a large difference between the target, planned, vaccination coverage
and the effective coverage obtained at the end of the campaign (see also~\citep{vaccinationFrance1}).

The 2009/2010 worldwide H1N1 epidemic spread through 213 countries and has been attributed $18\ \!\! 156$ deaths by 15/06/2010 (when epidemic was declared over by the WHO). In France
the first cases appeared in May 2009 and
 $1334$ severe forms were declared out of, approximately, $7.7$ to $14.7$ Millions 
people infected. The vaccination campaign in France was costly (around $500$M EUR, although cost estimation vary) for a
low efficiency of $8\%$ coverage (to be compared with $24\%$ in the US or
$74\%$ in Canada) although the target was set to $75\%$. During the French vaccination campaign 
some undesired neurological side-effects of the adjuvant present in the vaccines were under debate and known to the public.

The campaign was designed in waves (people were called to the vaccine centers according to their risk status) and all vaccination centers were not $100\%$ operational at once. Moreover the government mobilized additional medical personnel (military, medicine interns,...)
latter during the epidemic propagation. The joint combination of these effects is that the maximum vaccination capacity was non-constant increasing from zero to some peak obtained by the end of $2009$.

\subsection{Data sources} \label{subsec:datasources}

The information concerning the vaccination capacities were available in the general news but no quantitative estimation is, up to our knowledge, available. However the actual cumulative vaccination curve was reported by~\citet{couverturefrance} (up to 
the maximum coverage of $7.9\%$). 
The vaccine effectiveness coefficient was estimated by~\citet{efficacicevaccinFrance}
to be $71.9\%$. We set $U$ in the 
model~\eqref{sys:SIRU} to fit this data, see Figure~\ref{fig:fitvac}.

\begin{figure} 
\includegraphics[width=0.45\textwidth]{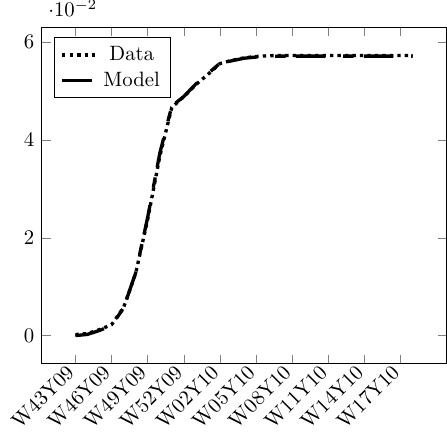}
\includegraphics[width=0.45\textwidth]{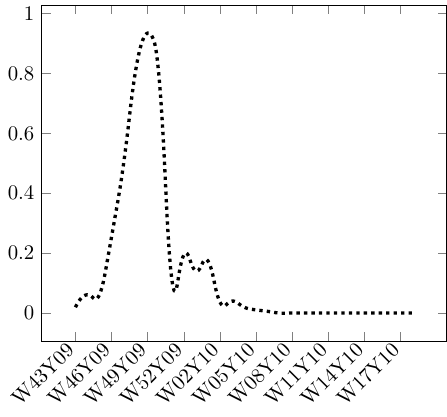}
\caption{
{\bf Left:} Fit of the cumulative vaccination in percents weighted by the effectiveness coefficient. The curve $U$ (labeled "Model") 
and the curve in~\citet{couverturefrance} (labeled "Data") cannot be distinguished. 
{\bf Right:} Instantaneous vaccination $dU$, weighted by the effectiveness coefficient.}
\label{fig:fitvac}
\end{figure}

In what concerns the size of the Infected class, the French "Sentinel" network (cf.~\cite{sentiweb}) reports the estimation of the number of infections per week starting from the beginning of the epidemic. 
Using the same data~\citet{grippeAdonnees1} compared the epidemic with seasonal epidemic dynamics. Finally~\citet{seroconversiongrippeA} 
measure seroconversion and obtain results on the number of people immunized at the end of the epidemics. 
All this data was used to estimate the curve $\ifd(t)$ in the model~\eqref{sys:SIRU}. The data obtained by the Sentinel network takes into account the number of consultations in medical offices. However some infected persons are asymptomatic (but contribute to the propagation) and thus only some percentage of the infected individuals will consult a medical doctor. This percentage is difficult to evaluate; we follow the~\citet{planningassumptions}
who estimated this rate at $50\%$; thus to estimate the number of infections we set $\ifd(t)$ two times larger than the curve reported by the Sentinel network. 

The propagation parameters $\beta$ and $\gamma$
were chosen consistent with ranges from the literature 
(see~\citet{pandemieH1N1donnes2}) although large confidence intervals were present, see for instance~\citep{grippeAdonnees2}.
We took  ${\mathcal R}_0= \beta/ \gamma = 1.35$.
We fit $\gamma$ in a very limited range, $\susinit$ in the range $0.8$ to $0.95$ (compatible with 
estimation concerning already immunized persons) and $\ifdinit$. The parameters were optimized in order to reproduce the cumulative number of infections $\int_{0^-}^\infty \beta \sus(t) \ifd(t)dt$ and the peak of the epidemic. 
We obtained $\gamma = 365/3.2$, $\susinit=0.84$ and $\ifdinit =2 \times 10^{-6}$.
The overall fit obtained is described in 
Figures~\ref{fig:fitvac} and~\ref{fig:fitinf}. Given the large uncertainties surrounding the propagation parameters and the number of infections the fit is considered very satisfactory.
Given the short time of propagation of the epidemic we took $\dis=0$.


\begin{figure} 
\includegraphics[width=0.45\textwidth]{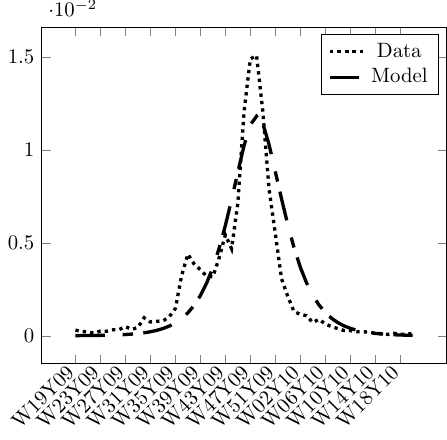}
\includegraphics[width=0.45\textwidth]{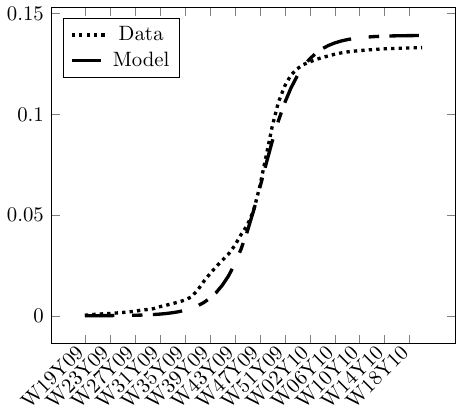}
\caption{Comparison between the Infected class  as obtained from the Sentinel network and the one from the model in~\eqref{sys:SIRU}. 
{\bf Left:} Infected class as function of time. The peak of the epidemic is well reproduced. {\bf Right:} cumulative number of Infections. The overall number of infection (at the end of the epidemic) is well reproduced. 
}
\label{fig:fitinf}
\end{figure}

\subsection{Methods}

Once the overall dynamics~\eqref{sys:SIRU} is given, there is still the quotient $\rv / \ri$ to be estimated. Such an endeavor is very difficult because this is depending on the individuals' perception and very few data is available on the heterogeneity of this parameter within the population. Rather than using very uncertain data we preferred to see what the model says about the quotient $\rv / \ri$. 

On the vaccination intensity Figure~\ref{fig:fitvac} (left) one can note that vaccination had a sharp increase up to week $50-51$ of $2009$ 
followed by a sharp decrease. The decrease was not due to the capacity of vaccination centers. 
We interpret it as reflecting a perception of a low infection risk coupled with high vaccination risk, i.e., a high 
quotient $\rv / \ri$. We can compute the precise value of the quotient $\rv / \ri$ compatible with our model. In 
Figure~\ref{fig:risklevels} we plot three risk indicators: 
first of them is $\frac{\vpI(\infty)- \vpI(t)}{1- \vpI(t)}$, 
the second $\frac{\NI(\sus(t), \ifd(t))}{\sus(t)}$; the third is the quotient between the number of infections from $t$ to $\infty$ divided by 
the number of susceptible people in $t$ (which is the initial value $\susinit$ minus  
the total infections and vaccinations between $0$ and $t$).
Note that this last criterion is model free and can be computed from 
the vaccination coverage report and the Sentinel network data (weighted by the asymptomatic infected individuals rate).

We looked therefore at the level of the three criterions that were attained at the time when vaccination decreased. According to our model this
level is an indication of the quotient $r_V/r_I$.

\subsection{Results}
An individual that stops vaccinating at the peak of the curve in Figure~\ref{fig:fitvac} (left) will have $r_V/r_I$ in the range
 $5\% - 10\%$. This is a huge value meaning that $1$ out of $10$ vaccinated people will develop side-effects as severe as the influenza A H1N1 itself. 
Such situations 
reflect probably a communication failure around the epidemic and the vaccines (infection risk, severity, vaccine side effects, need for a mass vaccination campaign, ...) rather than specific medical data about the risks involved. See also~\citep{determinantsvaccination} for a discussion on the determinants of individual vaccination and~\citep{Basu02122008}
for a related discussion concerning the controversial Human papillomavirus (HPV) vaccines.

Note also that vaccination did not stop completely at week $50-51$ of $2009$ but continued at detectable levels up to week $10$ of $2010$. An individual that stops vaccinating at week $10/2010$  time has $r_V/r_I < 1\%$ (probably much less as the model cannot be more precise with available data). This is a more common accepted figure.

Therefore at least two groups with very heterogeneous perceptions of the vaccination risk were present: a first group very worried about vaccine risk and another group less pessimistic.

We want to stress again that these results are highly dependent on the quality of the model and should be interpreted with care. Moreover the robustness of the results with respect to the input parameters $\gamma$, $\beta$, $dU$ have also to be assessed in practice.

\begin{figure} 
\sidecaption
\includegraphics[width=0.6\textwidth,height=0.4\textheight]{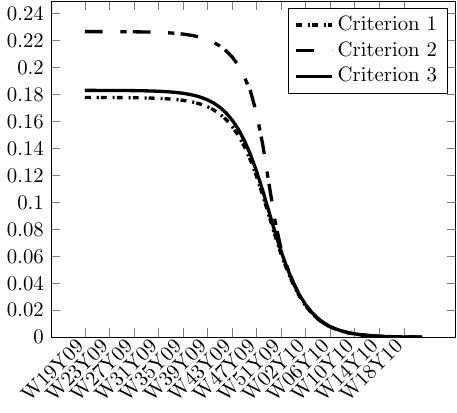}
\caption{Using the dynamics in~\eqref{sys:SIRU}, fitted with our data, we plot three criterions:
$\frac{\vpI(\infty)- \vpI(t)}{1- \vpI(t)}$,  $\frac{\NI(\sus(t), \ifd(t))}{\sus(t)}$ and the quotient between the infections in $[t,\infty[$ and the susceptibles in $t$ (see text).
By the time the first group of people stopped vaccination (week $50-51$ of $2009$) there was still
$5-10\%$ chance for an individual to contract the Influenza A. The individual that do not vaccinate at this point estimates the side effects of the vaccine to occur with frequency of about $5-10\%$, which is considered a very large and pessimistic value. Individuals that stopped last (week $10$ of $2010$) had this figure down to less than $1\%$ (probably much less given data uncertainties). 
Although the precise values may depend on the parameters we noticed that two distinct groups were found for a wide domain of parameters (compatible with data and with $dU$).
}
\label{fig:risklevels}
\end{figure}

\section{Discussion and conclusions} \label{sec:discussionconclusion}

In the context of individual vaccination (as opposed to global, compulsory vaccination) we develop an equilibrium model that allows to quantify the relationship between the individual perception of vaccine side-effects and of the epidemic morbidity.

The first important result is that, even when individuals are only driven by self interest, such an equilibrium exists. Secondly we are able to characterize the individual decision.
When there is no discounting
the individual will vaccinate or not depending on his estimations of the infection risk compared with the cost of the vaccine.
When discounting is taken into account the individual may choose to wait until epidemic risk rises and only then vaccinate.

The equilibrium is a Nash equilibrium with an infinity of players (as pioneered by~\citet{mfgcras1,mfgcras2}), which means in particular that it can have (and has) a non-null "price of anarchy". This means that self-interest, although stable from both individual-societal points of view, is not the best solution that can be obtained.

Finally, we apply the results to the Influenza A $2009$/$2010$ vaccination campaign in France and 
see that, under the assumptions of the model, 
it is hinted 
that at least two distinct groups existed, one very pessimistic about vaccine side-effects and the other rather optimistic.

This work has several limitations which open the way for future developments; the parameters in Section~\ref{sec:application} 
were calibrated in the following way: the reproduction number
$\gamma / \beta$ was taken in an interval centered around the parameters of seasonal influenza; on the contrary 
the initial number of Susceptibles ($\susinit$) and Infected ($\ifdinit$) was calibrated from data specific to this season. 
Parameters were optimized to fit the final number of Infected, Vaccinated and give a correct epidemic peak.
However it is not certain that only one set of parameters is compatible with data and the robustness has to be investigated, in particular for epidemics which are different from historical data.

In addition, 
our hypothesis that $\rv$ is known may be a limitation in general. 
Nevertheless, although vaccines are specific to each Influenza season, their risks are generally well understood and can be 
transferred from one season to another (but the vaccine immunity does not).
For the specific case of Influenza A (H1N1) 2009-10 season in France
it turns our that the vaccination arrived after several other European countries already started their vaccination programs; 
data from these earlier programs were largely discussed and even sparkled important controversies in France concerning the vaccine side-effects.
In general, when the risk $\rv$ is not known the model could be extended as in~\citet{waitandsee} 
where the authors discuss the learning of $\rv$ parameter during the Influenza A (H1N1) 2009-10 epidemic season in North America. 

Another limitation arises from the schedule of the vaccination campaign: 
people were assigned to risk groups and in principle, cannot vaccinate without a written invitation
received by mail (at a time corresponding to the risk group they belonged). However 
many centers were under-utilized by mid December 2009, see~\cite[page 26]{ccomptes} and on the other hand almost 2.7 Millons on-the-spot invitations were issued to people that presented at their initiative for vaccination, see~\cite[page 95]{rapportparlementaire}. We conclude that absence of invitations did not prevented, by itself, 
voluntary vaccination at the end of 2009.

\appendix

\section{On the societal and individual SIR model with vaccination} 
\label{appendix:sirequationmeasure} 

The mathematical description of the SIR model with vaccination has to take into account the possibility of instantaneous vaccination,
 which means that a non-negligible proportion of the total population can be vaccinated instantaneously at some time $t$. 
A first consequence is that $\sus(0)$ can be strictly less than $\susinit$ (when some vaccination occurs at time $0$).
Recall that $\vac(t)$ represents the  proportion of people vaccinated by the time $t$. If, for instance, nobody 
vaccinates before time $t^*=0.25$ years ($3$ months) and $30\%$ of the population vaccinates at time $t^*$ this means that $\vac(t)$ is discontinuous at this point (see Figure~\ref{fig:globalstrat} for an illustration). In particular it is not derivable and neither will be $\sus(t)$; as such it is not possible to use 
in Equation~\eqref{sys:SIRU} the derivative $d\sus(t)/dt$. This explains why the equation is only written in the differential form:
$d\sus(t) = - \beta \sus(t) \ifd(t) dt - dU(t)$. In this writing all objects have a well defined mathematical meaning: $dU(t)$ is a positive measure which, for our example will be the Dirac mass $0.3 \delta_{t^*}$. We refer to~\citet{bressanimpulse,rampazzo91,miller94measure,silvacontrolmeasure97} for the mathematical properties of the solutions to such evolution equations.
In the particular situation when $U(t)$ has a jump at $0$ we will have
$\susinit- \sus(0)=  U(0)- U(0^-)= U(0)$. 
\begin{figure} \sidecaption
\includegraphics[width=0.25\textwidth,height=0.15\textheight]{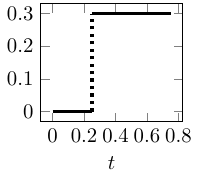}
\caption{Illustration of instantaneous vaccination of $30\%$ percent of the population at time $t^*=0.25$. The function 
$\vac(t)= \int_{0^-}^{t^*} dU(\tau)$ is plotted. It has a discontinuity at time $t^*$ and is equal to $H(\cdot-t^*)$ with $H(\cdot)$ the Heaviside function.}
\label{fig:globalstrat}
\end{figure}

In this dynamics all individuals are the same. Each individual is following a continuous time Markov dynamics jumping between the states 
Susceptible, Infected, Recovered and Vaccinated. 
Let $M_t$ be the state of one such individual at time $t$.
The time of the jump from the 
Infected to the Recovered class is a exponential random variable of mean $1/\gamma$. We write in terms of probabilities:
\beq
\P \left( M_{t + \Delta t} = Recovered \Big| M_{t} = Infected \right) = \gamma \Delta t + o(\Delta t).
\eeq
For a given individual in the Susceptible class at time $t$, the probability to be infected during the time interval $[t,t+\Delta t]$ is 
$\beta \ifd(t) \Delta t + o(\Delta t)$, therefore
\begin{equation} \label{eq:vpid1}
\P \left( M_{t + \Delta t}=Infected \Big| M_{t}=Susceptible \right)  = \beta I(t) \Delta t + o(\Delta t).
\end{equation}

On the other hand, 
denote by $\vpI(t)$ the probability of infection (in the absence of vaccination) during the time interval $[0,t]$. 
Then
\beqn
& \ & 
\P \left( M_{t + \Delta t}=Infected \Big| M_{t}=Susceptible \right)  =
\frac{ \P \left( M_{t + \Delta t}=Infected, M_{t}=Susceptible \right) }{ \P \left( M_{t}=Susceptible \right) }
\nonumber 
\\ & \ & \label{eq:vpid2}
=\frac{\vpI(t + \Delta t)- \vpI(t) }{1- \vpI(t)}
=\frac{1}{1- \vpI(t)}
\vpI'(t) \Delta t + o(\Delta t).
\eeqn

Passing to the limit $\Delta t \to 0$ in Equations~\eqref{eq:vpid1} and~\eqref{eq:vpid2}
we obtain:
\begin{equation}  \label{eq:vpI}
\vpI'(t) = \beta \ifd(t) (1 - \vpI(t)), \ \vpI(0^-) = 0.
\end{equation} 
therefore $\vpI$ is given by formula \eqref{eq:vpI_expli}.
Note that since $\ifd(t)$ is continuous $\vpI$ is differentiable everywhere.

In particular, for an individual in the Susceptible class at time $\tau$ that does not vaccinate any more from that time on,
the probability of infection after time $\tau$ is:
\begin{equation} \label{eq:vpi}
\frac{\vpI(\infty)- \vpI(\tau)}{1- \vpI(\tau)}.
\end{equation}

We can prove, by direct computations, that $dU=0$ on $[\tau, \infty[$ implies:
\begin{equation} \label{eq:vpi2}
\frac{\vpI(\infty)- \vpI(\tau)}{1- \vpI(\tau)} = \frac{\NI(\sus(\tau),\ifd(\tau))}{\sus(\tau)}.
\end{equation}

\begin{remark}
Since $\vpI(\infty) < 1$ the cost $\CIP{\Pi_\infty}{(\sus,\ifd)}{U}$ is not the limit of the costs $\CIP{\Pi_t}{\susinit, \ifdinit}{U}$ (for $t\to \infty$). On the contrary $\CIPd{\Pi_\infty}{(\sus,\ifd)}{U} = \lim_{t\to \infty} \CIPd{\Pi_t}{(\sus,\ifd)}{U}$. \\
\end{remark}

\section{Individual strategies} \label{appendix:individualstrat}

The simplest individual strategy is to vaccinate or not at some given time (provided he is still susceptible).  Such a strategy is called a 
{\it pure strategy}. However pure strategies do not always have good theoretical properties and 
in his Nobel award winning work John Forbes Nash proved that on the contrary, any finite game admits equilibrium if 
{\it mixed strategies} are allowed. A mixed strategy is a probability law on the set of all pure strategies. With the notations of 
Section~\ref{sec:themodel} the mixed strategy are probability laws on $[0,\infty]$ with special meaning of values $0$ (immediate vaccination)
and $\infty$ (no vaccination). The CDF function $\vpV(t)$ is such that $\vpV(0^-)=0$ (no vaccination before time $0$).
Note that $\vpV$ is not necessary a continuous function (the discussion is very much similar to the one in Appendix~\ref{appendix:sirequationmeasure}). In particular when the individual chooses the pure strategy $\Pi_{t^*}$ then
$\vpV$ is the Heaviside function $H(\cdot - t^*)$. See the illustration in Figure~\ref{fig:individualstrat}.
\begin{figure} \sidecaption
\includegraphics[width=0.35\textwidth,height=0.15\textheight]{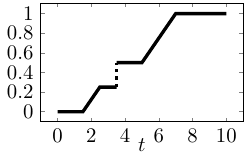}
\caption{
The individual cumulative probability of vaccination $\vpV(t)$ is an increasing, right continuous with left limits (c\`adl\`ag) function with $\vpV(0^-)=0$,  $\vpV(\infty)\le 1$.
}
\label{fig:individualstrat}
\end{figure}

It may also be necessary to impose some constraints. Suppose that global vaccination (at the society level) can 
only happen with the maximal rate of $u_{max}$ percent of population in a unit time. Then, with the notations in Equation~\eqref{sys:SIRU}:
$U(t+ \Delta t)-U(t)  \le u_{max} \Delta t$. Suppose now that all individuals want to vaccinate at the same time with the same $\vpV$ then the constraint above, coupled with~\eqref{eq:UGequilibre}, 
 implies 
{ that $\vpV$ is differentiable in any $t$ with $\sus(t) > 0$ and
denoting $f_V(t)=\frac{u_{max}}{\sus(t)}$:}
\begin{equation} \label{eq:constraint0}
{ \forall t \ge 0 \textrm{ with } \sus(t)>0:} \ 
\frac{d \vpV (t)}  {dt}  \le f_V(t) (1- \vpV(t)).
\end{equation}

Another interpretation of the constraint is the following: when the number of people that want to vaccinate exceeds the capacity of the vaccination centers people will have to wait. In this model all individuals are the same (that is, have the same characteristics therefore same strategies) then the probability for a given individual to obtain vaccination in a time interval $[t,t+\Delta t]$ is
$\frac{u_{max} \Delta t}{\sus(t)}$. The probability to
be not yet vaccinated at time $t$ but be vaccinated by time $t+\Delta t$ is on the one hand 
$\frac{\vpV (t+\Delta t) - \vpV (t)}{1-\vpV (t)}$ and on the other hand is 
$\left( \frac{u_{max} \Delta t}{\sus(t)} \right) + o(\Delta t)$; for $\Delta t \to 0$ 
we obtain the constraint 
\begin{equation} \label{eq:constraint1}
\frac{d \vpV (t)}{dt}  \le  \frac{u_{max} (1- \vpV(t)) }{\sus(t)}.
\end{equation}

\section{Proof of the properties of the individual optimal strategy: the undiscounted case}
\label{appendix:indoptstrat}

We set $\dis=0$.

\begin{theorem}[Case { $\um<\infty$}]
Let $U$ be a given societal policy in~\eqref{sys:SIRU}, increasing, $U(t)\le 1$, $\forall t\ge 0$
and 
{ $f_V(t)=\frac{u_{max}}{\sus(t)}$}.
Suppose that the set of admissible individual strategies is composed of all 
{ increasing functions $\vpV$ differentiable in any $t$ with $\sus(t) > 0$}
 such that:
\begin{equation}
\begin{cases}
{ \vpV(0^-)=0;}
\\
{ \vpV(\infty)\le 1};
\\
{ \forall t \ge 0 \textrm{ with } \sus(t)>0:} \ 
\frac{d \vpV (t)}{dt} \le f_V(t) (1- \vpV(t)).
\end{cases}
\label{eq:contrainte}
\end{equation}
Then the individual optimal strategy $\vpV^*$ that minimizes the cost in Equation~\eqref{eq:ci2} (with the system \eqref{sys:SIRU}) is:
\begin{enumerate}[a/]
\item \label{en:case1} If $\ri \le \rv $: $\vpV^*(t)=0$.
\item \label{en:case20} If $\ri > \rv $ and
$\vpI(\infty) \le \rv/\ri$ then $\vpV^*(t)=0$.
\item \label{en:case2} If $\ri > \rv $ 
and
$\vpI(\infty) > \rv/\ri$ then $\vpV^*(t)$ 
{ is the} solution of 
\beq 
\begin{cases}
{\vpV^*(0^-)=0, \vpV^*(\infty)\le 1, \ \vpV^* \textrm{ increasing }} 
\\
\forall t \in [0, \min\{Z_S, \si \}[: \
	\frac{d \vpV^* (t)}{dt} = f_V(t) (1- \vpV^*(t))
\\
\textrm{if } Z_S \le \si: \vpV^*(Z_S^-)=\vpV^*(\infty)=1 \\
\textrm{if } Z_S > \si: \vpV^*(\si^-)=\vpV^*(\infty)<1, \\
\end{cases}
	\label{eq:dvpV}
\eeq
with the notation
$Z_S=\inf_{\tau \ge 0}\{\tau | S(\tau) = 0 \}$.
\\
The parameter $\si$ is the unique solution of the equation: 
\begin{equation}
\frac{\vpI(\infty)- \vpI(\si)}{1- \vpI(\si)} = \frac{\rv}{\ri}.
\label{eq:zonevacbornee}
\end{equation} 
\end{enumerate}
\label{thm:ThetaIbornee}
\end{theorem}

\begin{proof} 
We use individual cost in the form in Equation~\eqref{eq:ci2}.
It may be noted that: 
\beq \label{eq:proof11}
 \rv - \ri \vpI(\infty) + (\ri - \rv) \vpI(t) = \rv(1-\vpI(t)) + (\rv - \ri) (\vpI(\infty) - \vpI(t)).
\eeq

\noindent {\bf Case~\ref{en:case1}}
If $\ri \le \rv$, since $\vpI$ is an increasing function, the quantity in~\eqref{eq:proof11} is the sum of positive terms. The minimum
attainable value is therefore zero and it obtained when $d \vpV(t) = 0 \ \forall t$. Or $\vpV(0^-) = 0$ thus $\vpV(t) = 0$ $ \forall t \ge 0$.

\noindent {\bf Cases~\ref{en:case20} and~\ref{en:case2}}
If $\ri \ge \rv$ let us compute
\beq
\frac{d}{d t} \left[ \rv - \ri \vpI(\infty) + (\ri - \rv) \vpI(t) \right] = 
(\ri - \rv) \frac{d \vpI (t)}{dt} \ge 0,
\eeq
(recall that $\vpI (t)$ is increasing) and moreover 
$ \rv - \ri \vpI(\infty) + (\ri - \rv) \vpI(0) =  \rv - \ri \vpI(\infty)$ is positive 
as soon as  $\vpI(\infty) \le \rv/\ri$.
To minimize the cost, vaccination should not occur when the term to integrate against $d\vpV(t)$ is positive 
therefore there is no vaccination if $\vpI(\infty) \le \rv/\ri$.

If $\vpI(\infty) > \rv/\ri$, vaccination occurs for all $t$ such that
$ \rv - \ri \vpI(\infty) + (\ri - \rv) \vpI(t) \le 0$, or equivalently 
$\frac{\vpI(\infty) - \vpI(t) }{1-\vpI(t)} \ge \frac{\rv}{\ri}$.

Moreover, 
\beq \label{eq:decroissancerisk}
\frac{d}{dt} \left[ \frac{\vpI(\infty) - \vpI(t) }{1-\vpI(t)} \right] = - \frac{[1- \vpI(\infty)]} {[1-\vpI(t)]^2} \frac{d \vpI(t)}{dt} < 0.
\eeq 
Then
$t \mapsto \frac{\vpI(\infty) - \vpI(t) }{1-\vpI(t)}$ is a decreasing, continuous 
function from $\vpI(\infty)$ to zero. To minimize the cost, the vaccination should appear 
right at the beginning and last until the boundary of the domain $\Omega$ is attained 
by the dynamics~\eqref{sys:SIRU}
or until time $\theta_I$, 
the unique solution of $ \frac{\vpI(\infty) - \vpI(\theta_I) }{1-\vpI(\theta_I)} = \frac{\rv}{\ri}$. 
Therefore in order to minimize the integral one has to set $d \vpV = 0$ on $[\theta_I, \infty[$ and 
{vaccinate}
on $[0, \theta_I[$ with maximal values coming first. Taking into account the constraint~\eqref{eq:contrainte} we obtain
{equation~\eqref{eq:dvpV}}
{
(the constraint $\um< \infty$ implies that if $Z_S \le \si$ then $\int_0^{Z_S} \frac{\um}{\sus(\tau)}d \tau = \infty$ 
thus $\vpV(Z_S^-)=1$).
}

\end{proof}

\begin{remark}
There is no individual vaccination if $\rv / \ri > 1- e^{-\beta/\gamma}$.
\end{remark}

The next results applies when the individual vaccination can be unbounded, i.e., $d\vpV$ can contain Dirac masses.
\begin{theorem}[Case {$\um=\infty$}]
The individual strategy $\vpV^*$ that minimizes the cost \eqref{eq:ci2} with the system \eqref{sys:SIRU} is:
\begin{enumerate}[a/]
\item \label{en:casea} if $\ri \vpI(\infty) < \rv$ then $\vpV^* \equiv 0$,
\item \label{en:caseb} if $\ri \vpI(\infty) > \rv$ then $d \vpV^* = \delta_0$,
\item \label{en:casec} if $\ri \vpI(\infty) = \rv$ then $d \vpV^* = \alpha \delta_0$ with $\alpha$ arbitrary in $[0,1]$.
\end{enumerate}
\label{thm:ThetaInonbornee}
\end{theorem}

\begin{proof}

\noindent {\bf Case~\ref{en:casea}} 
If $\int_{0^-}^{\infty} \rv - \ri \vpI(\infty) + (\ri - \rv) \vpI(t) d\vpV(t)$ is positive, then the minimum is greater than $\ri \vpI(\infty)$. This value is attained only if $\int_{0^-}^{\infty} \rv - \ri \vpI(\infty) + (\ri - \rv) \vpI(t) d\vpV(t) = 0$ so $\vpV(t) = 0$ for all $t$.

\noindent {\bf Case~\ref{en:caseb}} 
The strict monotony of the integrand 
$\rv - \ri \vpI(\infty) + (\ri - \rv) \vpI(t)$ allows to write:
\begin{align*}
\int_{0^-}^{\infty}& \rv - \ri \vpI(\infty) + (\ri - \rv) \vpI(t) d\vpV(t)  \\
&\ge \int_{0^-}^{\infty} \left[\rv - \ri \vpI(\infty) + (\ri - \rv) \vpI(t)\right]_{t=0^-} d\vpV(t) \\
&= \left[ \rv - \ri \vpI(\infty) \right] \int_{0^-}^{\infty} d\vpV(t) 
\ge  \rv - \ri \vpI(\infty),
\end{align*}
where we used $\vpV(\infty)\le 1$ and $ \rv - \ri \vpI(\infty)  \le 0$.
This gives a lower bound for the minimum. The bound is attained when both inequalities become equalities 
that is $d \vpV(t)$ is the Dirac mass in $0$.

\noindent {\bf Case~\ref{en:casec}} 
The difference with the previous case is that the last term 
$\rv - \ri \vpI(\infty) $ is $0$. Therefore 
 the last inequality is always satisfied. We obtain the conclusion.
\end{proof}

\section{Equilibrium strategy: the undiscounted case} \label{appendix:equilibre}

We set $\dis=0$.

\begin{theorem}[Case {$\um<\infty$}]
\label{thm:equilibriumborne}
Consider 
 admissible individual strategies 
{as in Theorem~\ref{thm:ThetaIbornee}.}
Then:
\begin{enumerate}[a/]
\item \label{en:casethm3a}
If
$\rv \ge \ri$ or $\rv < \ri$ and $\ifdinit + \rv / \ri \susinit +(\gamma/\beta) \ln(1-\rv/\ri) \le 0$ then the unique equilibrium is $U=0=\vpV$.
\item \label{en:casethm3b}
Otherwise consider the dynamics $(Y_1^{\infty}, Y_2^{\infty})$ starting from $\susinit, \ifdinit$ with 
$dU= u_{max} dt$ at all times until $\sus(t) =~0$. 
Let $\se^{\um}(\susinit, \ifdinit)$ be the first time when this dynamics touches the ensemble of curves:
\begin{equation}
\left\{
(\sus,\ifd) \in \overline{\Omega} \left|
\eqfrontiere = 0 \right.
\right\} \cup
\left\{
(\sus,\ifd) \in \overline{\Omega} \Big|
\sus=0 
\right\}.
\end{equation} 
Then the unique equilibrium is:
\begin{eqnarray}
& \ & 	dU(t) = \um \mathds{1}_{[0, \se^{\um}(\susinit, \ifdinit)]}(t) dt, 
	\label{eq:UGbornee}
\\
& \ & 
{\frac{d\vpV(t)}{dt}} = 
(1-\vpV(t)) {f_V(t) }\mathds{1}_{[0, \se^{\um}(\susinit, \ifdinit)]}(t).
	\label{eq:vpVbornee}
\end{eqnarray}
\end{enumerate}
\end{theorem}

\begin{proof} 
\noindent {\bf Case~\ref{en:casethm3a}} If $\rv > \ri$ then any optimal individual strategy is $\vpV=0$ and the compatibility 
relation~\eqref{eq:UGequilibre} imply $U=0$. Therefore the only possible equilibrium in this case is $U=0=\vpV$. 
It is easy to see that this is effectively an equilibrium.

If $\rv \le \ri$ and $\NI(\susinit, \ifdinit) \le (\rv/\ri) \susinit$, 
from Equations~\eqref{eq:vpi2} and~\eqref{eq:decroissancerisk}
 it follows that 
an individual strategy that vaccinates cannot be optimal, thus $\vpV=0$. As before we obtain the unique equilibrium $U=0=\vpV$. 

\noindent {\bf Case~\ref{en:casethm3b}} 
Since $\NI(\susinit, \ifdinit) > \rv/\ri \susinit$
any optimal individual strategy must satisfy~\eqref{eq:dvpV}.
Therefore, from the compatibility
relation~\eqref{eq:UGequilibre} we obtain
 $dU(t) = \um \mathds{1}_{[0, \se]}(t)$ for some $\se \ge 0$. 
We have to find $\se$ such that 
$\se$ is solution of \eqref{eq:zonevacbornee} for $dU(t) = \um \mathds{1}_{[0, \se]}(t)$. This is a fixed point equation.

Let $(\sus,\ifd)$ be the solution of the system \eqref{sys:SIRU} with $dU(t) = \um \mathds{1}_{[0, \se]}(t)$. 
Since $dU=0$ on $[\se,\infty]$ by Equation~\eqref{eq:vpi2}, any $\se$ solution of~\eqref{eq:zonevacbornee} 
is also solution of
\begin{equation}
\NI(\sus(\se), \ifd(\se)) = \rv/\ri \sus(\se).
\label{eq:frontiereTheta}
\end{equation}

But $(\sus, \ifd)= (Y_1^{\infty}, Y_2^{\infty})$ for $t \le \theta$, thus: 
\begin{equation}
\frac{\NI(\sus(\se), \ifd(\se))}{\sus(\se)} = \frac{\NI(Y_1^{\infty}(\se), Y_2^{\infty}(\se))}{Y_1^{\infty}(\se)}.
\end{equation}

Therefore, any $\theta$ that represents an equilibrium is also solution of the equation: 
\begin{equation}
\frac{\NI(Y_1^{\infty}(\se), Y_2^{\infty}(\se))}{Y_1^{\infty}(\se)} = \frac{\rv}{\ri}.
\label{eq:frontInf}
\end{equation}

But,
\begin{align*}
\frac{d}{dt} \left[ \frac{\NI(Y_1^{\infty}(t), Y_2^{\infty}(t))}{Y_1^{\infty}(t)}  \right]
&= \frac{1}{(Y_1^{\infty}(t))^2} 
\left[ \left(\frac{\partial \NI}{\partial \sus} \frac{d Y_1^{\infty}(t)}{d t} + \frac{\partial \NI}{\partial \ifd} \frac{d Y_2^{\infty}(t)}{dt} \right) Y_1^{\infty}(t) - \NI \frac{d Y_1^{\infty}(t)}{d t} \right]  \\
&= \frac{(\NI - Y_1^{\infty}(t))(\NI \um + \beta Y_1^{\infty}(t) Y_2^{\infty}(t)(\NI - Y_1^{\infty}(t) + \gamma/\beta)}{(Y_1^{\infty}(t))^2 (1+\beta/\gamma(\NI -Y_1^{\infty}(t)))} < 0,
\end{align*}
because $\NI < \sus$ and $\NI > \sus - \gamma/ \beta$ (see \citet{Laguzet2015180} for details and the expressions of the partial derivative of $\NI$).

Thus if \eqref{eq:frontInf} has a solution then this solution is unique. Therefore in any case at most one equilibrium exists.
Under the hypothesis $\NI(\susinit, \ifdinit) < \frac{r_V}{r_I} \susinit$ Equation~\eqref{eq:frontInf} does have a solution 
so $\se$ is also solution of Equation \eqref{eq:frontiereTheta} and this is the equilibrium.
\end{proof}

\begin{theorem}[Case {$\um=\infty$}] 
The equilibrium is:
\begin{enumerate}[a/]
\item \label{en:caseue1}
If $\rv \ge \ri$ or $\rv < \ri$ and
$\eqfrontiereinit \le 0$ then the unique equilibrium is $dU=0$ and $d\vpV= 0$ (no vaccination).
\item \label{en:caseue2}
If  $\ifdinit + (\gamma/\beta) \ln(1- \rv / \ri) > 0 $  then the unique
equilibrium is $dU=\delta_0 \susinit$ and $d\vpV= \delta_0$ (total instantaneous vaccination).
\item \label{en:caseue3}
Otherwise the unique equilibrium is $dU= \alpha^* \delta_0$ and $d\vpV = (\alpha^*/\susinit) \delta_0$ (partial instantaneous vaccination)
with 
$\alpha^* = \rv((\ifdinit + \rv/\ri \susinit + \gamma/\beta)\ln(1-\rv/\ri))/\ri $.
\end{enumerate}
\end{theorem}

\begin{proof}

Previous results indicate that in the unbounded case the individual optimal strategies are of the form
$d \vpV=\eta \delta_0$ thus $dU=\eta \susinit \delta_0$. Moreover, let $\alpha=\eta \susinit$ and recall that
$\sus(0)= \sus(0^-)- \alpha= \susinit-\alpha$; moreover:
\begin{align*}
\frac{d}{d \alpha} \left[ \frac{\NI(\susinit-\alpha, \ifdinit)}{\susinit-\alpha}  \right]
&= \frac{1}{(\susinit-\alpha)^2} 
\left[ \frac{\partial \xi}{\partial \sus} (-1) (\susinit-\alpha) - \xi (-1) \right]  \\
&= \frac{\NI (\NI - \susinit+\alpha)}{\left( \susinit - \alpha\right)^2 \left( \gamma/\beta + \NI - \susinit+ \alpha\right)} < 0,
\end{align*}
so the function $\alpha \rightarrow \frac{\NI(\susinit-\alpha, \ifdinit)}{\susinit-\alpha}$ is decreasing. Recall that $\NI(\susinit - \alpha, \ifdinit) / (\susinit - \alpha) = \rv/\ri$ is the same as $\ifdinit + \rv/\ri (\susinit - \alpha) + \gamma/ \beta \ln(1-\rv / \ri) = 0$. 
Similar arguments as in the proof of the Theorem~\ref{thm:equilibriumborne} apply and allow to reach the conclusion.
\end{proof}

\begin{proposition}
The vaccination region of the (OS) strategy contains
the vaccination region of the (EIS) strategy.
\label{ppt:emboitement_region}
\end{proposition}

\begin{proof}
In the (OS) strategy, the vaccination only stops 
if $\partial_{\sus} \NI \le \frac{\rv}{\ri}$ (see \citet{Laguzet2015180} for details).
Or  $\partial_{\sus} \NI = \frac{\NI}{\sus} \frac{1}{1+ \frac{\beta}{\gamma} (\NI - \sus)}$ and $\NI - \sus \le 0$ so $1+ 
\frac{\beta}{\gamma} (\NI - \sus) \le 1$. Then  $\frac{\NI}{\sus} \le \partial_{\sus} \NI \le \frac{\rv}{\ri}$ and the conclusion follows.
  \end{proof}

\section{Relation between global and individual cost.} \label{appendix:cout_moyen}

\begin{lemma}
Let $\vpV$ and $U$ satisfy Equation~\eqref{eq:UGequilibre}. Then
the individual cost is the average of the global cost, that is:
\begin{equation}
\frac{1}{\susinit} \CG{\susinit, \ifdinit}{U} = \CI{\vpV}{\susinit, \ifdinit}{U}.
\label{eq:moyenne}
\end{equation}
\label{lem:moyenne}
\end{lemma}

\begin{proof}
We can write:
\begin{equation}
\int_{0^-}^{\infty} \ri (1- \vpV(t)) d \vpI(t) = \int_{0^-}^{\infty} \ri (1-\vpV(t))(1-\vpI(t)) \left[- \frac{d \sus(t)}{\sus(t)} - \frac{d U(t)}{\sus(t)} \right] 
\label{eq:demPart1}
\end{equation}
because from~\eqref{sys:SIRU} for $\sus(t) \neq 0$: $\frac{d \vpI(t)}{1 - \vpI(t)} = \beta \ifd(t) =- \frac{d \sus(t)}{\sus(t)} - \frac{d U(t)}{\sus(t)}$.

Furthermore~\eqref{eq:UGequilibre} implies:
\begin{equation}
\int_{0^-}^{\infty} \rv (1-\vpI(t)) d \vpV(t) = \int_{0^-}^{\infty} \rv (1-\vpV(t))(1-\vpI(t)) \left[ \frac{dU(t)}{\sus(t)}\right].
\label{eq:demPart2}
\end{equation}
By summing the Equations \eqref{eq:demPart1} and \eqref{eq:demPart2} we get:
\begin{align}
\CI{\vpV}{\susinit, \ifdinit}{U} &= \int_{0^-}^{\infty} (1- \vpV(t))(1-\vpI(t)) \left[ -\ri \frac{d \sus(t)}{\sus(t)} -\ri \frac{dU(t)}{\sus(t)} + \rv \frac{dU(t)}{\sus(t)} \right] 
\nonumber
\\
	&= \int_{0^-}^{\infty} \frac{\sus(t)}{\susinit} \left[ -\ri \frac{d \sus(t)}{\sus(t)} -\ri \frac{dU(t)}{\sus(t)} + \rv \frac{dU(t)}{\sus(t)} \right]
\nonumber
 \\
	&= \frac{1}{\susinit} \int_{0^-}^{\infty}  \left[ -\ri d \sus(t) -\ri dU(t) + \rv dU(t) \right],
\end{align}
where we used $(1-\vpV(t))(1-\vpI(t)) = \sus(t)/\susinit$.
Using Equation \eqref{eq:vpI} and the definition of $\CG{\susinit, \ifdinit}{U}$ in Equation \eqref{eq:CG}, we obtain the result.
 \end{proof}

\begin{cor}
The average cost per person with the (OS) strategy 
is lower than the average cost per person with the 
(EIS) strategy.
\label{cor:cout_moyen}
\end{cor}

\begin{proof}
Denote by $U^G$ the (OG) strategy and $\vpV^G$ its individual counterpart. Also denote 
by  
$\vpV^I$ the individual strategy in the (EIS) equilibrium and $U^I$ its global counterpart.
 By the definition of the optimality of  $U^G$:
$\CG{\susinit,\ifdinit}{U^G} \le \CG{\susinit,\ifdinit}{U^I}$. From Lemma~\ref{lem:moyenne} after simplification by $\susinit$ we obtain:
$\CI{\vpV^G}{\susinit,\ifdinit}{U^G} \le \CI{\vpV^G}{\susinit,\ifdinit}{U^I}$ which is the conclusion.
  \end{proof}

\section{Some properties of the discounted number of infected} \label{appendix:zetad}

In this Section and in Appendix~\ref{appendix:equilibre_discount}
we consider $\dis >0$ and $u_{max}=\infty$.
Define $\Ietoile = (\rv \dis)/((\ri - \rv) \beta)$ and
$\NId ( \susinit, \ifdinit ) = \int_{0^-}^{\infty} e^{-\dis t} \beta \sus(t) \ifd(t) dt$ where
$(\sus(t),\ifd(t))$ is a non-vaccinating dynamics starting in $( \susinit, \ifdinit )$.
Denote also $\levelzd = \{ (\sus,\ifd) \in \Omega | \NId(\sus,\ifd) = (r_V/r_I) \sus  \}$.

\begin{lemma} \label{lemma:zetad}
\begin{enumerate}
\item \label{lemma:zetaditem1} The point $(0,\Ietoile)$ is always below the curve 
$\levelzd$. In particular either $\levelzd$ intersects the line $\ifd=\Ietoile$ in a point $(\Setoile,\Ietoile) \in \Omega$ or 
$\levelzd$ is completely above the line $\ifd=\Ietoile$.
\item \label{lemma:zetaditem2}
When  $\levelzd \cap \{(\sus,\ifd) \in \Omega|\ifd = \Ietoile\} \neq \emptyset$ , the 
non-vaccination dynamics starting from $(\sus,\ifd) \in \levelzd$ is entering the
domain $\{ (\sus,\ifd) \in \Omega | \NId(\sus,\ifd) \le (r_V/r_I) \sus \}$
if and only if  $\ifd \ge \Ietoile $.
\end{enumerate}
\end{lemma}
\begin{proof}
\noindent {\bf Item \ref{lemma:zetaditem1}:}
It is enough to prove that $\lim_{\epsilon \to 0} \NId(\epsilon,\Ietoile)/ \epsilon \le \rv / \ri$.
Let $(\sus(t),\ifd(t))$ be the evolution of the system without vaccination starting from the point $(\epsilon,\Ietoile)$. 
From $\ifd'(t) = (\beta \sus(t)-\gamma) \ifd(t)$ we obtain 

$\ifd(t) = \Ietoile e^{\int_{0^-}^t (\beta \sus(u) - \gamma) du} \in 
[\Ietoile e^{- \gamma t}, \Ietoile e^{(\beta \epsilon - \gamma) t}]$ and from
$\sus'(t) = (-\beta \ifd(t)) \sus(t)$ we obtain 
$\sus(t) = \epsilon e^{\int_{0^-}^t - \beta \ifd(u) du} \in 
[
\epsilon e^{\int_{0^-}^t - \beta \Ietoile e^{(\beta \epsilon-\gamma) u } du} , 
\epsilon e^{\int_{0^-}^t - \beta \Ietoile e^{-\gamma u } du} 
]
$. For $\epsilon \to 0$:
\beqn
& \ & 
\lim_{\epsilon \to 0}\frac{\NId(\epsilon,\Ietoile)}{\epsilon} = 
\int_{0^-}^\infty e^{-\dis t} \beta \Ietoile e^{-\gamma t } e^{-\beta \Ietoile \int_{0^-}^t e^{-\gamma u}du}dt
\nonumber \\ & \ & 
= \int_{0^-}^\infty - e^{-\dis t} \left(  e^{-\beta \Ietoile \int_{0^-}^t e^{-\gamma u} du}  \right)'
= 1 - \int_{0^-}^\infty \dis e^{-\dis t -\beta \Ietoile \int_{0^-}^t e^{-\gamma u}du} dt 
\nonumber \\ & \ & 
\le  1 - \int_{0^-}^\infty \dis e^{-\dis t -\beta \Ietoile t } dt = r_V/r_I, 
\eeqn
where, in the last inequality, we used that $\int_{0^-}^t e^{-\gamma u} du \le t$.

\noindent {\bf Item \ref{lemma:zetaditem2}:}
It is enough to show that  the tangent 
in $(\Setoile,\Ietoile)$ to  $\levelzd$
coincides with the direction $(-\beta \Setoile \Ietoile, \beta \Setoile \Ietoile - \gamma \Ietoile)$ (the dynamics of the system
without vaccination).
It is standard to prove (see \citet{Laguzet2015180}) that $\NId$ satisfies the following equation:
\begin{equation}
\partial_\sus \NId(\sus,\ifd) (-\beta \sus \ifd) +  \partial_\ifd \NId(\sus,\ifd) (\beta \sus \ifd-\gamma \ifd)
- \dis \NId(\sus,\ifd) +\beta \sus \ifd = 0.
\label{eq:hjbzetab} 
\end{equation}
On the other hand the normal to the curve $\levelzd$ 
in $(\Setoile,\Ietoile)$
is $(\partial_\sus \NId(\Setoile,\Ietoile)- \rv/\ri,\partial_\ifd \NId(\Setoile,\Ietoile))$. A simple computation shows that the condition 
\begin{equation}
(\partial_\sus \NId(\Setoile,\Ietoile)- \rv/\ri,\partial_\ifd \NId(\Setoile,\Ietoile)) \perp 
(-\beta \Setoile \Ietoile, \beta \Setoile \Ietoile - \gamma \Ietoile)
\end{equation}
reduces to~\eqref{eq:hjbzetab} when $\ifd=\Ietoile$.
\end{proof}

\section{Individual and equilibrium strategy with a discount factor} \label{appendix:equilibre_discount}

Let $g$ be the following function:
\begin{equation}
g(t) =  \ri (\phid(t) - \phid (\infty) ) + \rv e^{- \dis t} (1-\vpI(t)).
\label{eq:fctg}
\end{equation}

The individual cost functional can then be written as:
\begin{equation}
\CId{\vpV}{\susinit, \idfinit}{U} = \ri \phid(\infty) + \int_{0^-}^{\infty} g(t) d\vpV(t).
\label{eq:cid22}
\end{equation}

 Since $\ifd(t)$ is continuous with continuous derivative the same is true for  $g$ and 
\begin{equation}
g'(t) =  e^{-\dis t} (1-\vpI) \Big[ (\ri - \rv) \beta \ifd - \rv \dis \Big].
\label{eq:derive_fctg}
\end{equation}
Therefore the function g is increasing when $ \ifd < \Ietoile$ and decreasing otherwise. On the other hand $\ifd$ 
 decreases to zero at $\infty$ and may remain superior to $\Ietoile$ on a bounded time interval. As such, depending on $(\susinit,\ifdinit)$ and $U$,
 the following possible behaviors can occur:

- $g(t)$ is decreasing from $g(0) >0$ to $g(\infty)=0$

- $g(t)$ is decreasing from $g(0)$ to some value $g(t_1)$, increases from $g(t_1)$ to $g(t_2)$ and then
decreases from $g(t_2)$ to $g(\infty)=0$.

The minimum of $\CId{\vpV}{}{}$ is realized as following:
\begin{enumerate}[A/]
\item  If $g(t) >0$ for all $t$ then  $\vpV=0$ (never vaccinate).

\item \label{item:strategywait} If $\inf_{t \ge 0} g(t)<0$ then the optimum is realized when $d \vpV$ is a Dirac mass 
placed at the (unique) time $\tau_*$ such that $g(\tau_*)=\inf_{t \ge 0} g(t)$. In particular $\tau_* < \infty$.

\item Otherwise there exists an unique $\tau_*$ such that 
$g(\tau_*)=0$ and the optimal strategies are $d\vpV= \alpha \delta_{\tau_*}$ ($\delta_{\tau_*}$ being the Dirac mass in $\tau_*$),
$\alpha \in [0,1]$ arbitrary.

\end{enumerate}

\begin{remark}
For the individual strategy described in item~\ref{item:strategywait} 
it is optimal to delay vaccination to a latter time. 
Such a strategy is never encountered when $\dis=0$, where vaccination occurs either at $t=0$ or never. 
\end{remark}

Finally, as in Equation~\eqref{eq:vpi2} we can prove that if $dU=0$ on $[\tau, \infty[$, for any $t\ge \tau$
the value $g(t)$ is  positive / negative / null  if and only if 
$\NId(\sus(t),\ifd(t))$ is less / larger / equal to  $(\rv/\ri) \sus(t)$. 

Consider the following notations:
\begin{enumerate}[I)]
\item When, with the notations of Lemma~\ref{lemma:zetad}, 
a point $(\Setoile,\Ietoile) \in \levelzd$ exists let $\Setoiled$ be the unique solution in $[\gamma/\beta,1]$ of the equation:
$x - (\gamma/\beta) ln(x) = \Ietoile + \Setoile - (\gamma/\beta) ln(\Setoile)$.

Then the domain $\Omega$ is divided in three subdomains:
$\Omega_i= \{ (\sus,\ifd) \in \Omega | \NId(\sus,\ifd) > (\rv/\ri) \sus, \ifd > \Ietoile \} $, 
$\Omega_d=\{ (\sus,\ifd) \in \Omega | \ifd < \Ietoile, \ifd+\sus - (\gamma/\beta) ln(\sus) >  \Ietoile + \Setoile - (\gamma/\beta) ln(\Setoile) \}$, 
$\Omega_n = \Omega \setminus \overline{\Omega_i \cup \Omega_d } $.

\item When $\levelzd$ is above the line $\ifd=\Ietoile$, 
$\Omega_i= \{ (\sus,\ifd) \in \Omega | \NId(\sus,\ifd) > (\rv/\ri) \sus\} $, 
$\Omega_n= \{ (\sus,\ifd) \in \Omega | \NId(\sus,\ifd) < (\rv/\ri) \sus\} $.
\end{enumerate}

\begin{theorem}
Let $(\susinit,\ifdinit) \in \Omega$; the unique equilibrium of the individual-societal vaccination is the following:
\begin{enumerate}[a/]
\item \label{en:caseue1dn}
If $(\susinit, \ifdinit) \in \Omega_n$ then $dU=0$ and $d\vpV= 0$ (no vaccination).
\item \label{en:caseue1di}
If $(\susinit, \ifdinit) \in \Omega_i$ let $\alpha \in [0,1]$ be the maximum value such that 
$( \susinit - \alpha, \ifdinit ) \in \overline{\Omega_i}$.
Then  
$dU= \alpha \delta_0$ and $d\vpV= (\alpha/\susinit) \delta_0$ (partial or total instantaneous vaccination).
\item \label{en:caseue1dd}
If $(\susinit, \ifdinit) \in \Omega_d$ let $\tau_*$ be the time at which
the system (without vaccination) evolving from $(\susinit, \ifdinit)$ reaches the line
$\ifd=\Ietoile$.
Then $dU=(\sus(\tau_*^-)-\Setoile) \delta_{\tau_*}$ and 
$d\vpV= (\sus(\tau_*^-)/\Setoile-1) \delta_{\tau_*}$ (vaccination after waiting the time $\tau_*$).
\end{enumerate}
\end{theorem}

\begin{proof}
We will consider only the more involved situation when 
a point $(\Setoile,\Ietoile) \in \levelzd$ exists.

Recall that the evolution of the system without vaccination satisfies 
$ \ifd+\sus - (\gamma/\beta) ln(\sus) = cst$. Therefore the frontier of $\Omega_d$ and $\Omega_n$ is a trajectory of the system without vaccination.
Let us consider the global vaccination strategy $U$ given in this Theorem. This strategy 
does not vaccinate in domain $\Omega_n$, 
vaccinates (instantaneously) in domain $\Omega_i$ and  
when the evolution starts in domain $\Omega_d$ 
it waits to reach the line $\ifd=\Ietoile$; 
at that time it vaccinates until reaching the point
$(\Setoile,\Ietoile)$. This dynamics is illustrated in Figure~\ref{fig:regions_discount}.
 In order to prove that this is effectively an equilibrium
we still have to prove that the best individual policies $\vpV$ are coherent with $U$.

But the properties of the function $g(t)$ show that $g(t)$ is positive in  $\Omega_n$: therefore 
for a starting point $(\susinit,\ifdinit)\in \Omega_n$ the best individual strategy is to never vaccinate.

When the starting point $(\susinit,\ifdinit) \in \Omega_i$, the instantaneous vaccination makes it arrive 
at time $0$ on the boundary $\Omega_n$ and $\Omega_i$; therefore $g(0^-)=g(0)=0$. Optimal individual strategies
are $\eta \delta_0$ with $\eta \in [0,1]$; among those only one is coherent with $U$ (the one described in item~\ref{en:caseue1di}).

Finally, when the starting point $(\susinit,\ifdinit) \in \Omega_d$, the free (non-vaccination) evolution 
makes it arrive at some point $(\sus(\tau_*^-), \Ietoile)$. Then at time $\tau_*$ it reaches the point
$(\Setoile,\Ietoile)$ thus $g(\tau_*)= g(\tau_*^-)=0$. But, before $\tau_*$ the coordinate $\ifd$ was inferior to $\Ietoile$ 
thus $g(t)$ was decreasing to zero during this time. Then  
 all $\eta \delta_{\tau_*}$, $\eta\in [0,1]$ will be optimal strategies. Among those, only one value of $\eta$ is compatible with $U$.
Thus the strategies proposed in this Theorem are equilibrium strategies.
Uniqueness is proved as in Theorems~\ref{thm:ThetaIbornee} and \ref{thm:ThetaInonbornee}.
\end{proof}

\begin{remark}
For $\dis > 0$ and $\um < \infty$ we were not able to find an analytic expression for the domains
$\Omega_i$, $\Omega_n$ and $\Omega_d$. The individuals,
being aware of the shortage of vaccines, will wait in line in advance to be vaccinated by the optimal time; it is possible to obtain the corresponding Hamilton-Jacobi-Bellmann equilibrium equation but the equation has to 
be solved numerically.
\end{remark}


\end{document}